# Using Deep Learning to Identify Artificial Satellite Trails in Multi-band Photometric Astronomical Images

Hua-Jian Yu[1], Jia-Lei Zheng[2] and Yuan Fang[3*]

[1] First author: South-Western Institute for Astronomy Research, Yunnan University, Kunming 650500, P.R.China; *huajianyu@mail.ynu.edu.cn*

[2] Second author: South-Western Institute for Astronomy Research, Yunnan University, Kunming 650500, P.R.China; *zhengjialei@stu.ynu.edu.cn*

[3] Corresponding author: South-Western Institute for Astronomy Research, Yunnan University, Kunming 650500, P.R.China; *fangyuan@ynu.edu.cn*

Abstract This research addresses the growing challenge of artificial satellite trail interference in ground-based astronomical observations by developing an efficient deep learning identification method. With the proliferation of satellite constellations in low Earth orbit, accurate detection of satellite trails has become crucial for preserving astronomical data quality. Using multi-band photometric survey observational data from the *Multi−channel Photometric Survey Telescope* (Mephisto) of Yunnan University , we constructed a specialized dataset of astronomical images containing satellite trails. We propose a novel ASA-U-Net model that integrates atrous spatial pyramid pooling with channel attention mechanisms into the UNet architecture to effectively capture sparse satellite trail features that traditional semantic segmentation models often miss during downsampling. The model was implemented and validated on actual telescope data, demonstrating superior performance in end-to-end detection and marking of satellite trails compared to traditional methods. This approach significantly improves data processing precision without requiring manual parameter adjustments, making it suitable for processing massive nightly survey data and enhancing the quality of astronomical data products.

# 1 INTRODUCTION

Modern astronomy is an observational empirical science that aims to answer major scientific questions through massive data acquired from planned sky surveys. With advancing telescope technology, astronomical science has entered the big data era, with breakthroughs in both primary mirror diameter and detector performance parameters like quantum efficiency and pixel scale.

Ground-based survey systems like the Sloan Digital Sky Survey(York et al. 2000), Pan-STARRS (Chambers et al. 2019), and the upcoming Vera Rubin Observatory's Legacy Survey of Space and Time (Andreoni et al. 2024) will continuously collect photometric data for billions of celestial objects. However, the high sensitivity of modern detectors makes them more susceptible to interference from cosmic rays and artificial satellite trails. As commercial space companies like SpaceX and OneWeb accelerate deployment of low Earth orbit satellite constellations, the density of satellite trails in astronomical images continues to grow, becoming commonplace.

Time-domain surveys face increasing data processing demands - LSST, for instance, aims to cover the entire sky every three days over ten years, processing approximately 20 TB of raw data nightly. In this context of high data volume and real-time processing requirements, rapidly and accurately identifying non-celestial signals in astronomical images is critical for providing high-quality data products for survey science.

Artificial satellites orbiting Earth along fixed paths can interfere with astronomical observations due to their operational altitude and reflective properties. The tens of thousands of planned satellites will inevitably disrupt professional astronomy, with apparent magnitudes reaching 3 to 7 (Mroz et al. 2022). These satellites´ leave bright streaks in astronomical images, severely interfering with celestial signals. (Walker et al. 2020) identified several astronomical fields particularly vulnerable to large LEO satellite constellations, including optical and infrared wide-field imaging and transient observations. Charge Coupled Device (CCD) sensors used in astronomy require long exposures for effective observation of specific targets (Howell 2006), making them particularly susceptible to satellite trail interference. These trails appear as linear high-brightness features that can cover target source regions, distorting photometric measurements and significantly impacting image quality.

With the rapid increase in artificial satellites, ground-based astronomical telescopes require more advanced data processing methods and observation scheduling strategies to mitigate their impact. According to relevant survey reports by (Walker et al. 2020), current practical methods to reduce the impact of Earthorbiting satellites on optical astronomical observations include: First, adopting special materials to reduce satellite reflectivity of solar radiation. Second, utilizing accurate satellite ephemerides to avoid specific observational pointings or suspend exposures when satellites traverse the observed celestial region. Third, removing or masking satellite trails to eliminate their influence on photometric images. However, these methods may interfere with timely acquisition and accurate processing of scientific data, and compared to image processing algorithms, they entail higher operational costs and longer implementation cycles.

In terms of software processing methodologies, researchers have proposed various techniques to identify and eliminate artificial satellite trails in single-exposure astronomical images. These methods can be broadly categorized into source detection methods, signal template fitting approaches, computer vision techniques, and deep learning algorithms (Stoppa et al. 2024). Regarding source detection methods, SExtractor software (Bertin & Arnouts 1996) is commonly employed for source detection and identification of elongated objects, a methodology that has been implemented in the Palomar Transient Factory project. However, this approach demonstrates poor performance under low signal-to-noise ratio (SNR) conditions and frequently yields high false-positive rates (Waszczak et al. 2017).The signal template fitting approach aligns predefined signal morphologies with image data and calculates the weighted sum of pixels along the signal profile to quantify the degree of correspondence. This matched filtering methodology was proposed by (Turin 1960) in the 1960s and applied by (Dawson et al. 2016) for detecting asteroid and satellite trails in astronomical images, assuming uncorrelated data noise and constant point spread function (PSF), employing maximum likelihood detection. While enhancing accuracy, this method is computationally intensive, requiring matched filtering for each region of interest, and a single matched filter cannot comprehensively accommodate all practical scenarios, thereby affecting detection efficacy.

Computer vision techniques provide an alternative category of detection tools, demonstrating effectiveness across various scenarios. The most universally applied algorithm for identifying artificial satellite trails involves Hough transform processing of astronomical images (Hassanein et al. 2015). Its fundamental principle posits that satellite trails can be conceptualized as ensembles of randomly distributed points approximately aligned along a straight line; the Hough transform maps points from the original image to corresponding lines in Hough space, where line intersections correspond to the geometric parameters of satellite trails in the original astronomical image space, thereby facilitating satellite trail identification. However, for complex image scenes, this method may misidentify data noise or dense point sources as linear features. Additionally, different stellar fields and image backgrounds may necessitate parameter adjustments to achieve optimal performance. (Nir et al. 2018) proposed detecting satellite trails by cross-correlating astronomical images with PSF-processed signal templates to extract trail morphologies. To efficiently accomplish this, the Radon transform is employed to perform cross-correlation of segment positions and angles, integrating pixel values along all potential trails in the image. Direct computation of the Radon transform requires substantial computational resources; searching an $N \times N$ sized image necessitates scanning all linear streak starting positions and angles, increasing computational demands and accumulating image noise, thereby compromising result accuracy. (Stark et al. 2022) introduced an improved variant of the standard Radon transform. This enhanced transformation methodology, through integration with median filtering techniques, minimizes the influence of non-linear feature data such as stars and galaxies, significantly augmenting the detection capability for low-luminosity satellite trails. However, it exhibits sensitivity to median filtering parameter selection, where suboptimal parameters may adversely impact overall detection performance. (Kollo et al. 2023) proposed a novel automated threshold determination method that dynamically adjusts thresholds for diverse regions within astronomical images, reducing false detections. However, due to the requirement for multiple threshold adjustments, it imposes high computational demands, consuming substantial resources when processing high-resolution astronomical images, rendering deployment challenging on data backend systems with real-time requirements.

Deep learning methodologies have demonstrated superiority over traditional manual or rule-based programming approaches in numerous specific scenarios, achieving performance that sometimes surpasses human expert capabilities, presenting extensive application potential. Recent years have witnessed an increasing trend in deep learning applications within astronomical research (Meher & Panda 2021), indicating that deep learning methods and models have been widely implemented across numerous astronomical domains, proving highly effective for astronomical big data analysis and processing.

Deep learning offers novel opportunities for enhancing detection efficiency and accuracy through training convolutional neural networks (Lecun et al. 1998) on extensive image datasets. These methodologies can directly learn complex patterns and features from data, improving detection efficiency and reducing false-positive rates. (Paillassa et al. 2018) proposed utilizing a SegNet-like neural network architecture to detect all non-celestial signals affecting scientific observations in astronomical images. Following semantic segmentation tasks, this network architecture outputs a probability for each pixel, assigning probabilities of belonging to non-celestial signals or celestial targets. Since training and testing data were derived from specific telescope observations, the model's generalizability to practical observational images is relatively limited. (Paillassa et al. 2020) subsequently proposed employing the VGG-based classifier MaxiMask to perform image semantic segmentation tasks, constructing training samples by incorporating non-celestial signals into observational images. For satellite trails, they simulated trails using SkyMaker software to generate closely spaced stellar images of identical dimensions along linear paths. (Tiwari 2022) proposed utilizing generative adversarial networks to remove selected foreground objects from images and applying transfer learning to tasks employing convolutional neural networks for eliminating rain bands in specific weather scenarios. While satellite trails in astronomical images can be eliminated by modifying model parameters and providing appropriate training data, this method may compromise the quality of astronomical data proximate to the trails during the elimination process.

The methodologies referenced in the aforementioned literature are specifically designed for particular telescope data or extensive simulation datasets, characterized by high model complexity and training costs, rendering them impractical for direct application to observational data from Yunnan University's Mephisto Survey Telescope. There remains a need for a methodology with high real-time performance capability that can process increasingly complex astronomical image scenes and large-scale astronomical data. Building upon these previous advancements, our research employs the U-Net model (Ronneberger et al. 2015) from the deep learning semantic segmentation domain, integrated with Channel Attention Mechanisms(Hu et al. 2019) and Atrous Spatial Pyramid Pooling module(Chen et al. 2017), to detect and mark artificial satellite trails in astronomical images from the Mephisto Survey Telescope for subsequent photometric data analysis.

The organization of this paper is as follows: in Section ??, we provide a detailed description of the Mephisto telescope and the dataset samples used in this study; in Section 3, we describe the ASA-U-Net model and training strategy, as well as the metrics for evaluating model performance; in Section 4, we present the performance of the ASA-U-Net model after training and testing; finally, we summarize our work in Section 5. The catalog is available online.

## 2 DATA

This study uses data from the Mephisto telescope located in Yunnan Province, China. The telescope employs a three-channel spectroscopic architecture (Yuan et al. 2020) with an innovative design that transcends the technical limitations of traditional survey observations. This configuration enables simultaneous threeband observation capability, allowing the acquisition of *ugi/vrz* multi-band observational data in a single exposure, thereby obtaining real-time color information of celestial objects. Currently, the telescope continues to provide high-precision, high-quality observational data for astronomical research, playing a crucial role in frontier fields including galaxy cosmology, time-domain astronomy, and galactic archaeology.

In this section, we explain the data used to create ASA-U-Net and the dataset-building process. This involves selecting, preparing, and manually labeling images to ensure high-quality training data for our deep learning algorithm.

### 2.1 Mephisto Telescopes

The Mephisto telescope, with its 1.6-meter primary mirror aperture and a highresolution $9216 \times 9232$ and $6144 \times 6160$ pixel CCD, offers a wide field-of-view of 3.14 square degrees, sampled at $0.56''$/pixel. Compared to the single-band sequential observation limitations of existing survey projects, Mephisto's three-channel synchronous observation advantage overcomes the critical bottleneck of delayed color information in transient and variable source research. It features red ($i/z$ bands), yellow ($g/r$ bands), and blue ($u/v$ bands) channels equipped with three large mosaic CCD cameras, with a total pixel count exceeding 1 billion, comparable to LSST.

The telescope uses integration times ranging from 30s to 600s. Consequently, under these observational conditions, the trails of artificial satellites become more prominent, and the brightness of satellite trails varies across different observational wavelength bands. These trails leave streaks in survey photometric images that affect target observations; therefore, identifying and marking these satellite trails can significantly improve the quality of data processing. The following content, we provide survey observational data from three channels in different wavelength bands for explanation and demonstration.

*2.1.1 Red Channel Camera Data*

The Mephisto Survey Telescope's red channel camera operates in the $i$-band with a wavelength range of $755nm$ to $900nm$, and the $z$-band with a wavelength range of $900nm$ to $1050nm$. When imaging in these $i$ and $z$ bands, the telescope exhibits a Fringing effect, as shown in Fig. 1. The Fringing effect occurs when the CCD is influenced by atmospheric emission lines in the $i$ and $z$ bands during observation; light reflects within the CCD and creates interference patterns, forming stripe-like structures. In astronomical research, this effect impacts the photometric measurements of celestial objects and the quality of image subtraction. Unlike astronomical images obtained through optical imaging in other wavelength bands, identifying artificial satellite trails in astronomical images with Fringing effects presents a particularly challenging task.

*2.1.2 Yellow Channel Camera Data*

The Mephisto Survey Telescope's yellow channel camera operates in the $g$-band with a wavelength range of $480nm$ to $580nm$, and the $r$-band with a wavelength range of $580nm$ to $680nm$. Since the peak radiation wavelengths of $F$, $G$, and $K$-type main sequence stars predominantly fall within the $g$ and $r$ band ranges, more stars can be detected with the same exposure time, resulting in significantly higher stellar density in these astronomical images compared to images in other bands, as shown in Fig. 2. When identifying artificial satellite trails in $g$ and $r$ band observational images, the interference from high stellar density must be excluded.

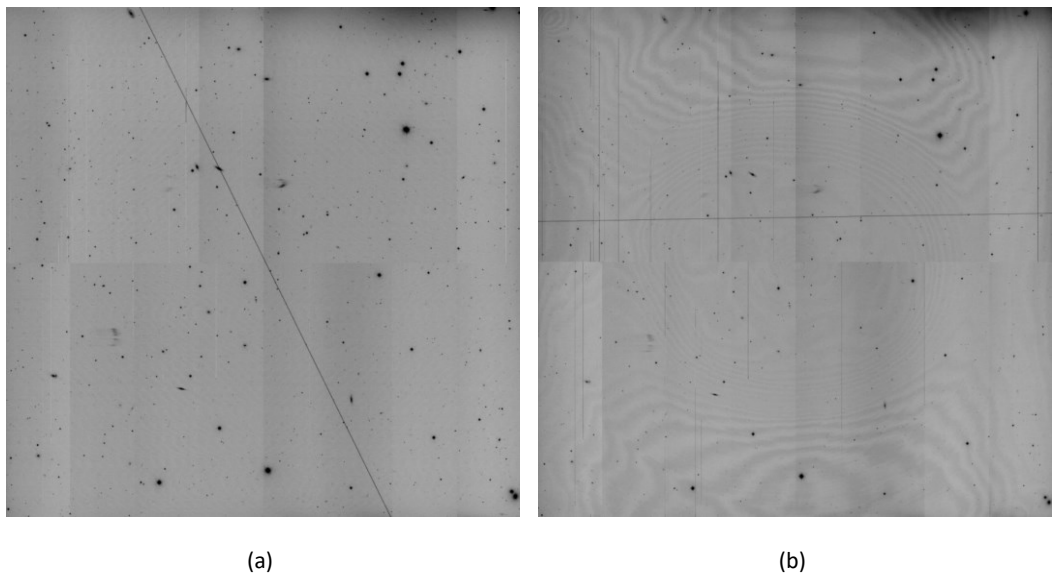

(a)          (b)

Fig.1: Artificial satellite trails captured in astronomical images with Fringing effects. (a) Artificial satellite trails photographed in the $i$-band. (b) Artificial satellite trails photographed in the $z$-band. We can observe that the Fringing effect in (a) is less prominent than in (b), yet both images exhibit artificial satellite trails traversing the field of view.

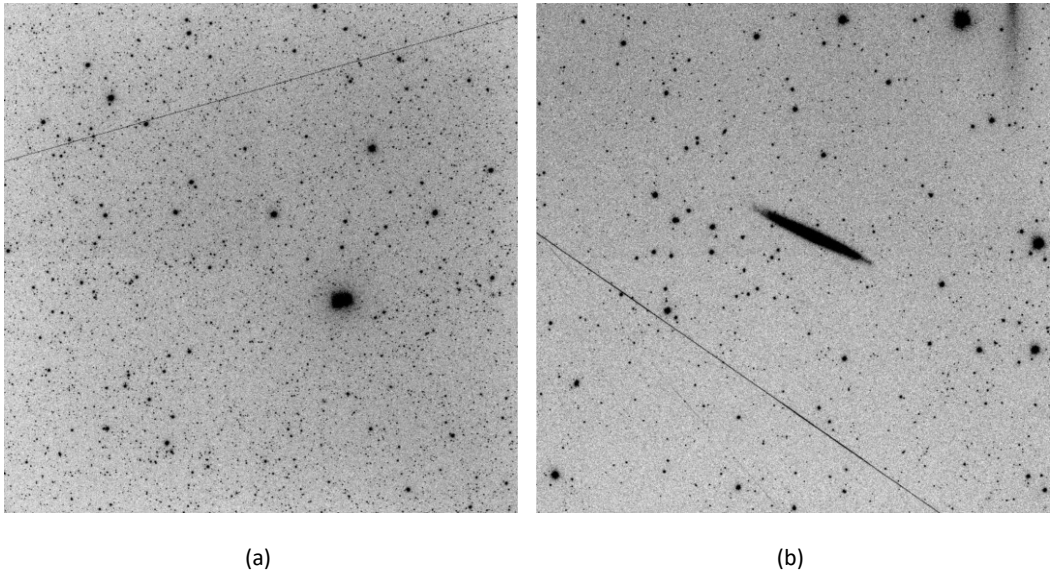

(a)　　　　　　　　　　　　　　　　　　　　(b)

Fig.2: The (a) and (b) show artificial satellite trails in dense stellar fields captured in the $g$-band and $r$-band, respectively.

*2.1.3 Blue Channel Camera Data*

The Mephisto Survey Telescope's yellow channel camera operates in the $u$-band with a wavelength range of $320nm$ to $365nm$, and the $v$-band with a wavelength range of $365nm$ to $405nm$. In astronomical observations, due to Earth's atmosphere having stronger absorption and scattering effects on this set of wavelengths compared to other longer wavelengths, images captured in the $u$-band and $v$-band have lower pixel values and appear dimmer, as shown in Fig. 3. Therefore, in $u$ and $v$ band images, the task of identifying artificial satellite trails transforms into one of detecting faint objects, which places higher demands on satellite trail detection algorithms.

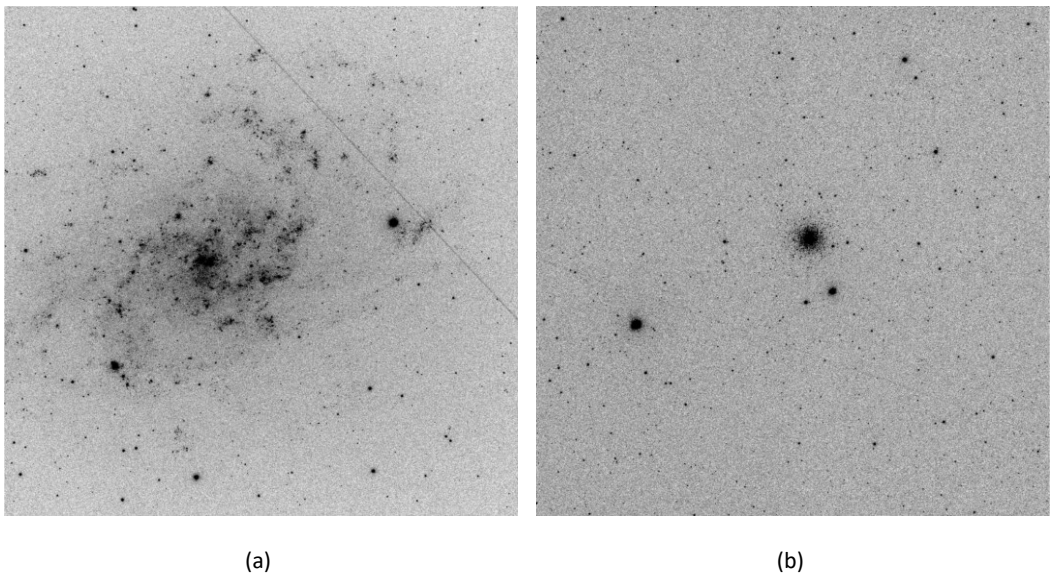

(a)　　　　　　　　　　　　　　　　　　　　(b)

Fig.3: The (a) and (b) show faint artificial satellite trails in $u$-band and $v$-band images, respectively.

*2.1.4 Comparison of Three-band Data*

In this section, we provide two sets of survey observation data from different bands, showcasing the *igu/zrv* three-band observational data simultaneously acquired during a single exposure. Currently, the red channel camera of the telescope has an image size of $9.2k \times 9.2k$ pixels, while the yellow channel and blue channel cameras have image sizes of $6.2k \times 6.2k$ pixels. Therefore, the images displayed here are drawn according to the proportional sizes of different channel cameras, as shown in Fig. 4 and Fig. 5. It can be observed that, following the survey observation plan, the three cameras simultaneously observed the same sky area with identical exposure times. Due to the different observation bands, the resulting images exhibit varying characteristics. Based on this, we categorize different datasets according to the different channel cameras.

2.2 Dataset preparation

Astronomical images often contain various non-scientific information, such as defective pixel regions caused by observational equipment flaws, captured satellite trails, star trailing, charge saturation bleeding, and more. This non-scientific information may interfere with subsequent image processing and scientific analysis, thus requiring preprocessing of image data.

The astronomical image data used in this study was downloaded from the Yunnan University Mephisto Survey Telescope database. The original image files obtained from telescope observations are in FITS (Flexible Image Transport System) format, which can store observational images from astronomical telescopes and related scientific data.

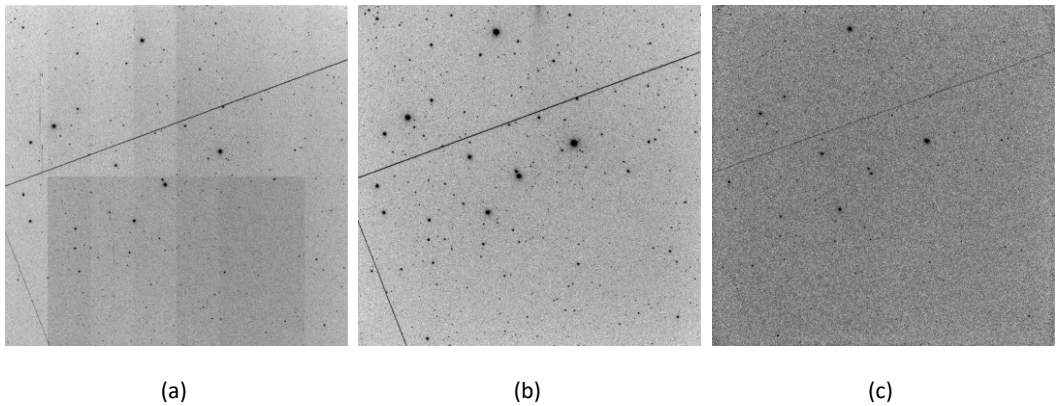

(a) (b) (c)

Fig.4: The (a), (b) and (c) represent observational image data from three different bands: the *i*-band, *g*-band, and *u*-band, respectively.

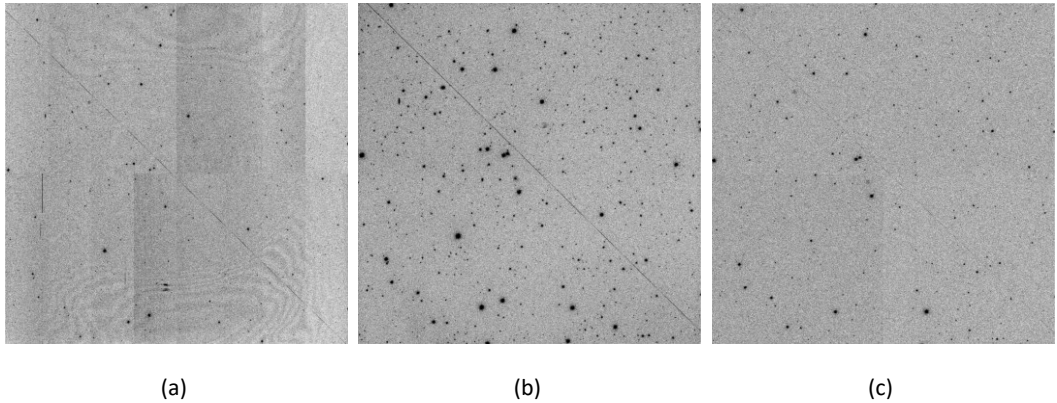

(a)                    (b)                    (c)

Fig.5: The (a), (b) and (c) represent observational image data from three different bands: the *z*-band, *r*band, and *v*-band, respectively.

Although FITS format files are widely used in astronomy, there are certain difficulties in directly using FITS format files for deep learning processing. This is because most neural network models and deep learning frameworks typically better support processing common PNG (Portable Network Graphics) or JPEG (Joint Photographic Experts Group) image formats. PNG format is a bitmap image format that uses lossless compression algorithms, preserving high-quality image details with relatively small file sizes, making it very suitable for scientific data visualization processing. To more conveniently utilize neural networks for astronomical image processing, we converted FITS format image files to PNG format image files, with the converted PNG format images being 8-bit integers.

Considering that the Mephisto Survey Telescope's observational data has image sizes of $6k \times 6k$ and $9k \times 9k$ pixels, totaling tens of millions of pixels, training on full images would impose a tremendous computational burden on equipment. More importantly, in astronomical image processing tasks based on deep learning, the features of observational targets often manifest in local regions. Cutting images into multiple smaller pictures can encourage deep learning models to focus more on learning these critical local feature patterns, thereby improving the model's feature extraction capability.

After downloading FITS files containing satellite trails from the Mephisto Survey Telescope database and converting them to PNG images, we further cut them into multiple 512×512 pixel images and selected those containing satellite trails. This segmentation strategy both ensures the integrity of local target features and significantly reduces hardware computational resource requirements. The model can greatly improve its prediction effectiveness by learning from multiple sample images containing satellite trails.

Our established satellite trail dataset includes two categories of objects: satellite trails and astronomical background, with the task objective being to predict the category of each pixel in the image. We used labelme software to manually annotate astronomical images and create label files. In the labelme software, points were manually plotted to determine the range of satellite trails within the images, generating json files with label information. Finally, binary label images were generated based on the json files. Fig. 6 displays image patches containing satellite trails from the Red, Yellow, and Blue Channel Cameras (*ugi/vrz* bands) and their corresponding ground truth-labels.

In this paper, we performed the same data augmentation operations after completing the annotations on the original astronomical images, ensuring that the newly generated images and labels after data augmentation align in pixel positions. We used the Albumentations (Buslaev et al. 2020) library in Python to perform image data augmentation operations. This library allows synchronous transformation of images and their corresponding labels, effectively expanding the dataset and thereby effectively enhancing data diversity to ensure the model's generalization ability. The data augmentation methods used in this dataset were horizontal and vertical flipping, which helps increase the diversity of positive samples, simulating satellite trails appearing in different directions, and improving the deep learning model's detection capability for satellite trails from various directions.

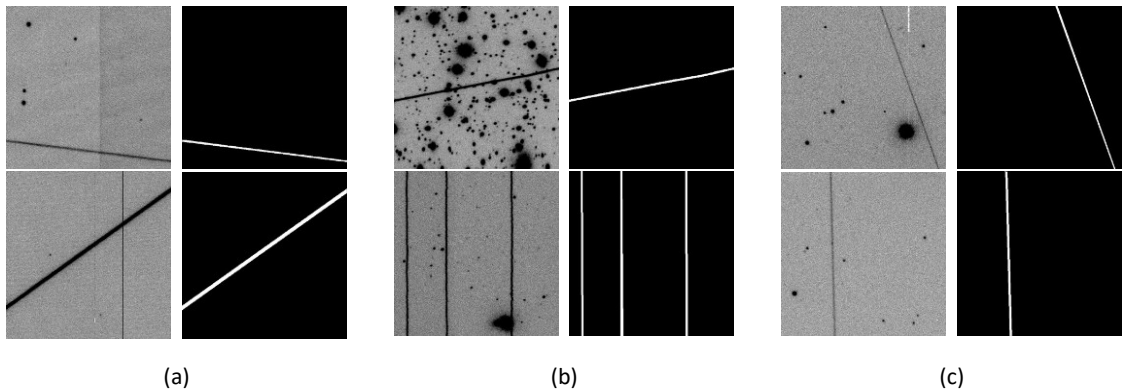

(a)　　　　　　　　　　(b)　　　　　　　　　　(c)

Fig.6: Examples of satellite trail and ground truth-labels in Mephisto images. The figure shows data content from three groups: (a) Image data from the Red Channel Camera. (b) Image data from the Yellow Channel Camera. (c) Image data from the Blue Channel Camera. For each group, the first column shows astronomical images containing satellite trails, and the second column shows the corresponding ground-truth labels.

## 3 STRUCTURE, TRAINING AND EVALUATION OF ASA-U-NET

Our research aims to accurately segment artificial satellite tracks from astronomical images, a task framed as a semantic segmentation problem where every pixel is classified as track or background. To this end, we developed a deep convolutional neural network based on the U-Net architecture, a model renowned in segmentation for its encoder-decoder structure and skip connections.

Nevertheless, standard U-Net models face unique challenges when applied to astronomical data. Unlike common images, astronomical frames contain distinct features: satellite tracks can be exceptionally faint due to high orbits or low reflectivity, making them difficult to distinguish from the background. This background is itself complex, populated by celestial objects of varying shapes and brightness that can be easily mistaken for tracks. In addition, the satellite tracks themselves vary significantly in their width, brightness, and continuity. Consequently, an effective model must possess the capability to handle all these disparate image properties simultaneously.

This section will provide a detailed introduction to the newly proposed ASA-U-Net model, which can be used for detecting satellite trails in astronomical images, including its architectural design and training strategies, and performance metrics.

## 3.1 The structure of ASA-U-Net

To overcome the challenges previously outlined, we have developed a customized modification of the standard U-Net (Ronneberger et al. 2015), which we call the Astro Satellite-Aware U-Net(ASA-U-Net). Our key enhancements focus on two main areas.

First, to improve the extraction of faint satellite track features amidst complex astronomical backgrounds, we have integrated Residual connections structure (He et al. 2015) and a Channel Attention Mechanism (Hu et al. 2019) within our fundamental convolutional blocks. This strategy strengthens the discriminative power of features while simultaneously suppressing irrelevant background noise.

Second, to effectively capture satellite tracks of diverse lengths and widths, we have incorporated a multi-scale Atrous Spatial Pyramid Pooling (ASPP) (Chen et al. 2017) module. This structure replaces the conventional convolutional layers at the U-Net bottleneck, enabling the model to perceive contextual information at multiple scales within a single, cohesive framework.

In essence, our model is built upon a U-Net skeleton, augmented by these specialized modules, with the goal of achieving precise and robust satellite track segmentation.The structure of the proposed ASA-U-Net model is shown in Fig 7. The following sections will provide a detailed analysis of each module's design and its specific function in addressing the difficulties of this astronomical task.

### 3.1.1 Satellite-Aware Convolutional Network Module

In the standard U-Net model, the core unit at each level consists of two consecutive 3 × 3 convolutional layers, which are responsible for extracting image features from the input data. However, the feature extraction capability of this standard double convolutional module is limited when processing astronomical images with low signal-to-noise ratios and complex backgrounds, often leading to the loss of critical target information. To address this issue, we have designed an Satellite-Aware Convolutional Network (SACN) Module. This module builds upon the traditional double convolution structure by incorporating a Residual
Connection and a Channel Attention mechanism in series, aiming to effectively enhance the target information.

The architecture and operational sequence of this module are illustrated in Fig 8.

Given an input feature map $X_{in} \in R^{C \times H \times W}$ to the module, where $C$, $H$, and $W$ represent the number of channels, height, and width of the image, respectively, the module first performs residual processing.The input $X_{in}$ passes through a first 3 × 3 convolution, Batch Normalization (BN), and a ReLU activation function, followed by a second 3 × 3 convolution and Batch Normalization. The output of this process is then added element-wise to the original, unprocessed input $X_{in}$, and a final ReLU activation is applied. The output of this residual learning stage, that $X_{res}$ can be represented by equation (1).

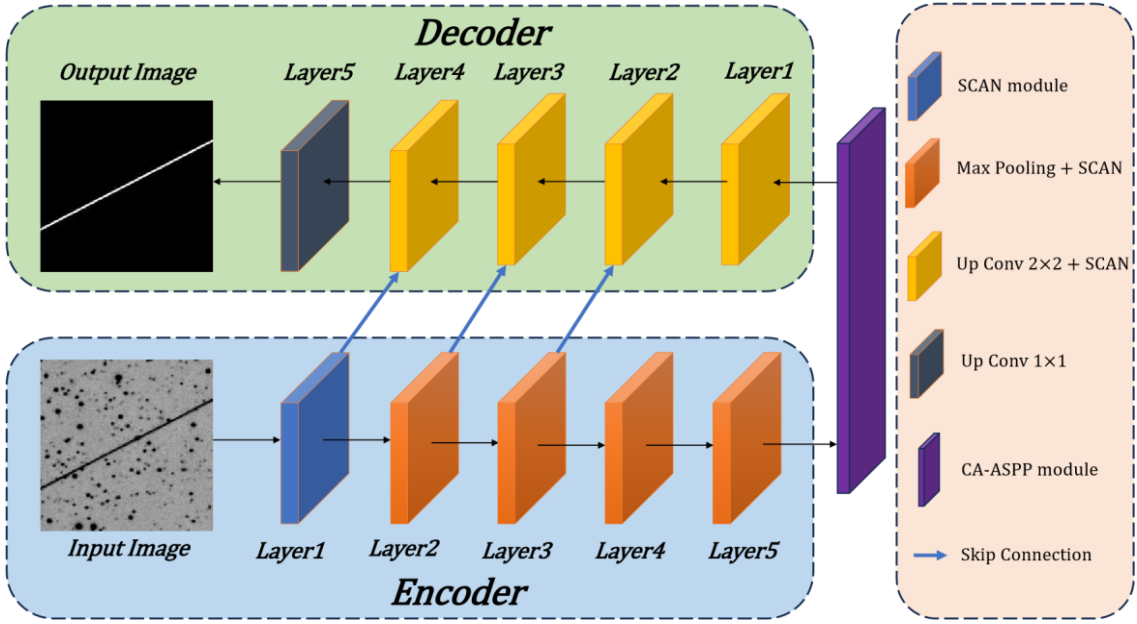

Fig.7: ASA-U-Net model's structure

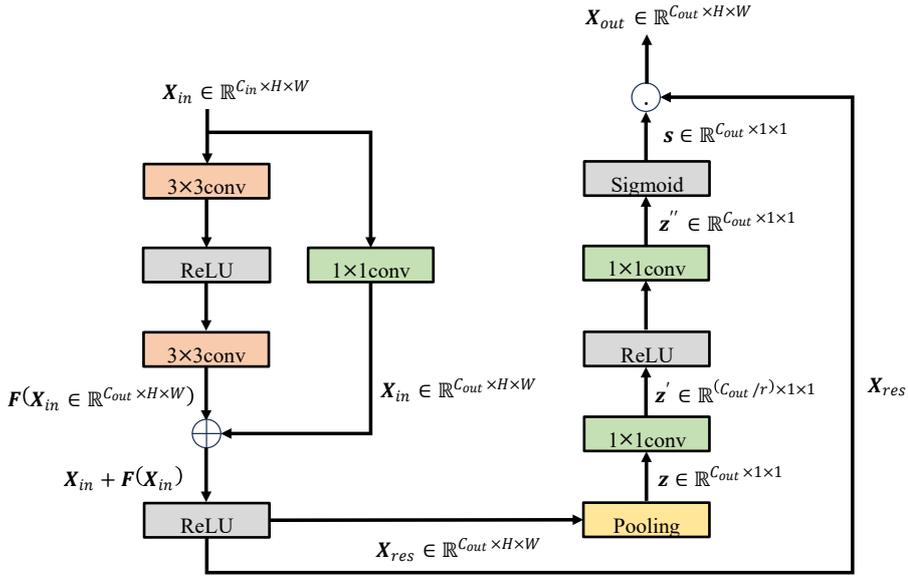

Fig.8: Structure of the Satellite-Aware Convolutional Network (SACN) Module. The input feature $X \in \mathbb{R}^{C \times H \times W}$ first passes through a residual convolutional block, producing the residual feature $X_{res}$ while preserving spatial resolution. Subsequently, the channel attention mechanism is applied to compute channelwise features and generate a weight vector $s \in \mathbb{R}^{C \times 1 \times 1}$. The weight is then broadcast and multiplied with the residual feature to obtain the final output $X_{out} \in \mathbb{R}^{C \times H \times W}$.

$$X_{res} = ReLU(BN(W_2 \otimes (ReLU(BN(W_1 \otimes X_{in}))))) + X_{in}. \tag{1}$$

Here, $\otimes$ denotes the convolution operation, and $W_1$ and $W_2$ represent the convolution kernels.

Through this residual structure design, we mitigate the common problem of gradient vanishing during the training of deep networks. More critically, it ensures that the original input $X_{in}$ can be directly connected to the module's output, effectively preserving the original astronomical image information even after multiple non-linear transformations. Subsequent modules can then further optimize the model parameters based on this preserved information.

After processing by convolutional layers, astronomical images produce multiple feature channels, with each channel holding varying levels of importance for different tasks. By incorporating a Channel Attention mechanism, the model can automatically evaluate and identify the feature channels that are most useful for recognizing artificial satellite trails while suppressing channels with irrelevant interference, thereby improving overall performance.

We feed the output from the previous stage, $X_{res}$, into the channel attention module. Through a Global Average Pooling (GAP) operation, the global information of each channel in $X_{res}$ is compressed into a single numerical value, forming a channel descriptor vector $z$. The calculation for the $c$-th element, $z_c$, of this vector is shown in equation (2).

$$z_c = \frac{1}{H \times W} \sum_{i=1}^{H} \sum_{j=1}^{W} X_{res}(i, j, c). \tag{2}$$

Here, $H$ and $W$ denote the height and width of the feature map, respectively, and $X_{res}(i,j,c)$ represents the value of the feature map $X_{res}$ at spatial location $(i,j,c)$ in channel $c$. Subsequently, the vector $z$ is fed into a 1 × 1 convolutional layer for dimensionality reduction to learn the non-linear inter-channel dependencies. The dimension of vector $z$ is reduced from $C$ to $C/r$ (wherer $r$ is the reduction ratio, set to 16 in our model) to decrease the model's computational complexity. After a ReLU activation, a second 1x1 convolutional layer restores the vector dimension to $C$, a Sigmoid activation function is used for normalization to generate the importance weight for each channel. The final operation, as shown in equation (3), involves the element-wise multiplication of the channel weight vector $s \in R^{C \times 1 \times 1}$ with the feature map $X_{res}$ via a broadcasting mechanism.

$$X_{out} = X_{res} \odot s. \tag{3}$$

Here, $\odot$ denotes the element-wise multiplication, Through this module, channels that are highly correlated with the features of artificial satellite trails are assigned higher weights and are thus enhanced, while channels related to celestial bodies or other image backgrounds are effectively suppressed. In summary, our Enhanced Double Convolutional Module first preserves original image information through a residual connection structure and then utilizes channel attention to significantly improve the model's ability to identify satellite trails in complex astronomical backgrounds.

*3.1.2 Atrous Spatial Pyramid Pooling with Channel Attention module*

In the segmentation task of identifying artificial satellite trails, another challenge lies in the significant variations in target scales. Satellites in low Earth orbit may appear as broad and bright streaks in astronomical images, whereas those in higher orbits or of smaller sizes may present as thin and faint trails. If the model only relies on convolutional kernels of fixed size, it may perform well on a particular scale but tends to overlook trails of other scales. To address this issue, we introduce the Atrous Spatial Pyramid Pooling (ASPP) module into the bottleneck layer of the model. The core idea of ASPP is to adopt a parallel branch structure that extracts multi-scale feature representations through convolutional kernels with different receptive fields. we design the CA-ASPP (Atrous Spatial Pyramid Pooling with Channel Attention) module, which integrates multi-scale feature extraction and channel re-weighting.

The ASPP module consists of five parallel branches that simultaneously process the input feature map $X \in R^{C \times H \times W}$ from the encoder stage. As illustrated in Fig 9, the branches include one standard $1 \times 1$ convolution, three atrous convolutions with different dilation rates, and one pooling branch. The outputs of these branches are concatenated along the channel dimension, and then compressed into a more compact yet efficient feature space via a $1 \times 1$ convolution, yielding the unified multi-scale representation $X_{aspp}$, as shown in equation (4), where $W_0$ is the weight of the 1×1 convolution, and $W_{r1}$, $W_{r2}$, $W_{r3}$ are the weights of dilated convolutions with dilation rates $r_1$, $r_2$, $r_3$, respectively. The operator $[\cdot]$ indicates concatenation along the channel dimension, $X_{Upsample}$ is the global context feature obtained by broadcasting the pooled vector back to the spatial size $H \times W$. Finally, $X_{aspp} \in R^{C_{out} \times H \times W}$ is the output feature of the ASPP module, $C_{out}$ is the number of channels obtained after concatenation. On top of this, we further incorporate a channel attention module to compute the weights of the most informative feature channels for satellite trails detection, while suppressing less relevant or distracting information.

$$X_\text{aspp} = W_{aspp} \otimes [W_0 \otimes X_{in}, W_{r1} \otimes X_{in}, W_{r2} \otimes X_{in}, \\ W_{r3} \otimes X_{in}, Upsample(GAP(X_{in}))], \tag{4}$$

3.2 Model training strategy

The hardware parameters for the experiments include an Intel Core i9-14900 CPU and an NVIDIA GeForce RTX 4090 24G GPU. All networks are implemented using Ubuntu 22.04 operating system, PyTorch deep learning framework, and Python programming language. During the training phase, we iterated a total of 100 epochs. The initial learning rate was set to 0.0001, and the optimizer used is AdamW (Loshchilov & Hutter 2019) . Learning rate decay is implemented using cosine annealing, and the weight decay regularization coefficient is 0.01.

Considering the thin, elongated geometric characteristics of satellite trails against astronomical image backgrounds, when the proportion of positive samples (satellite trails) in the classification task dataset is significantly lower than negative samples (astronomical background), the model tends to favor the class with more negative samples during training, resulting in poor prediction performance for positive sample classes. To overcome the sample imbalance problem, we employed the FocalLoss loss function (?), which is particularly suitable for semantic segmentation tasks in identifying satellite trails in images. Through a dynamic factor, it dynamically reduces the weight of easily distinguishable samples during training, thereby quickly focusing attention on difficult-to-distinguish samples. This approach improves the model's detection accuracy on class-imbalanced datasets.

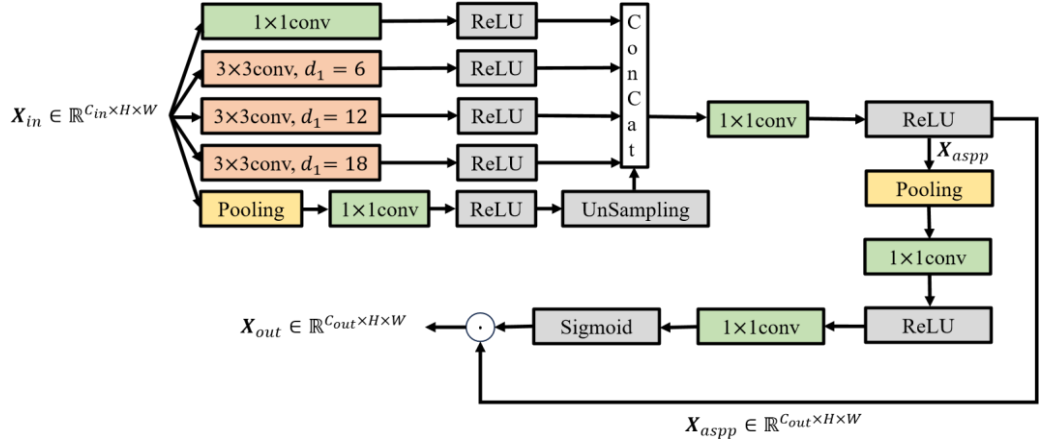

Fig.9: Structure of the Atrous Spatial Pyramid Pooling with Channel Attention module (CA-ASPP). The input feature map is processed by five parallel branches: a 1 × 1 convolution, three 3 × 3 atrous convolutions with different dilation rates (6, 12, and 18), and a global average pooling branch. The outputs of all branches are concatenated along the channel dimension, and then, a channel attention mechanism is applied to recalibrate the fused feature map by learning channel weights from global context.

Figure 10 illustrates the change in loss values during the training process of the ASA-U-Net model. The graph shows that as the training epochs increase, both the training and validation losses gradually decrease and stabilize. This indicates that the ASA-U-Net does not suffer from overfitting or underfitting issues.

3.3 Performance metrics

This section introduces metrics for evaluating object detection networks. This paper employs Mean Pixel Accuracy(MPA), Precision, Recall, Dice-Coefficient, and IoU as fundamental metrics for evaluating model performance, with their calculation formulas corresponding to such as equation (5), (6), (7) and (8), (9) below.

$$MPA = \frac{1}{k+1} \sum_{i=0}^{k} \frac{TP}{TP+FP}.$$
(5)

$$Precision = \frac{TP}{TP + FP}. \tag{6}$$

$$Recall = \frac{TP}{TP + FN}. \tag{7}$$

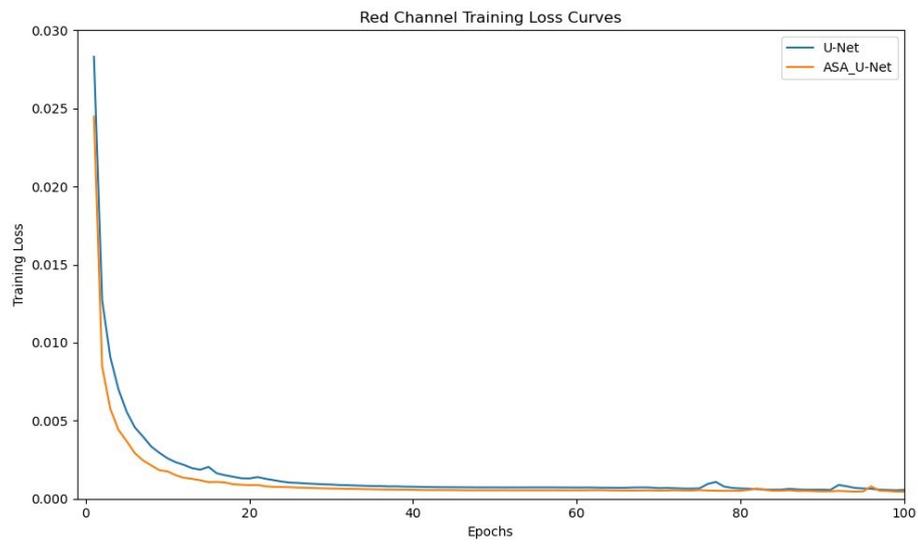

(a)

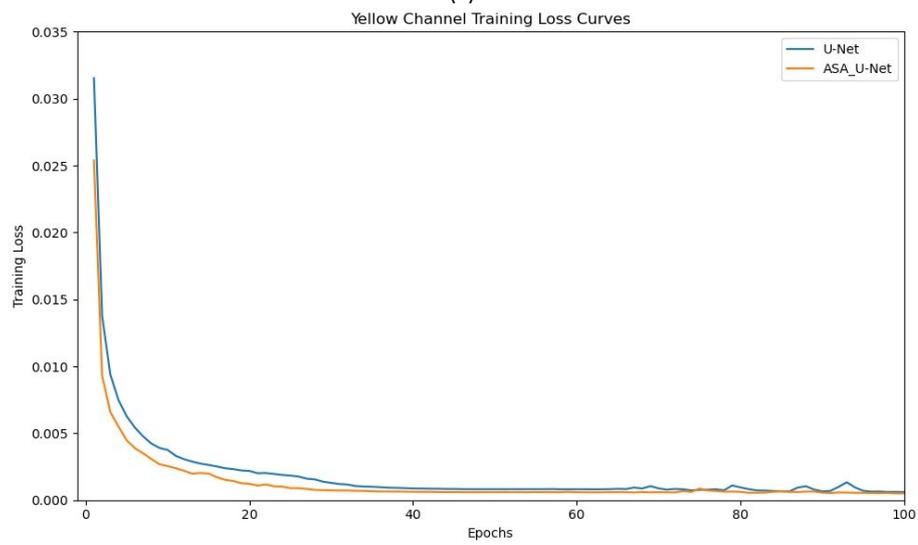

(b)

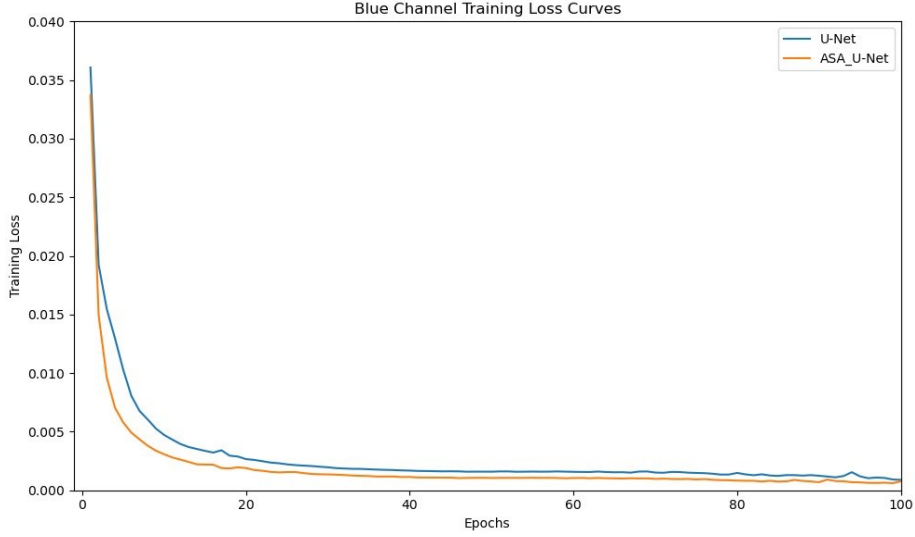

(c)

Fig.10: The (a), (b) and (c) represent ASA-U-Net model training loss curves in three different channels: the red channel, the yellow channel, and the blue channel, respectively.

$$Dice - Coefficient = \frac{2 \times TP}{2 \times TP + FN + FP}. \tag{8}$$

$$IoU = \frac{TP}{TP + FP + FN}. \tag{9}$$

Here, The Mean Pixel Accuracy (MPA) metric is employed to assess a model's overall classification performance(in this task, $k = 1$), especially on datasets with a significant class imbalance. It begins by independently calculating the per-class accuracy—the proportion of pixels correctly classified within each category—and subsequently computes the arithmetic mean of these values. This method circumvents the issue of the standard Overall Accuracy metric becoming artificially high due to a predominant background class. Precision represents the proportion of pixels that truly belong to a specific category among all pixels predicted as that category. Recall measures the proportion of successfully identified pixels of a true positive category among all pixels of that true positive category. The Dice coefficient is a metric commonly used to evaluate the similarity of two samples. It measures the degree of overlap between the model's predicted segmentation region and the ground truth region. The value of this metric ranges from 0 to 1, where a higher value indicates greater similarity. Intersection over Union (IoU) is a commonly used evaluation metric in semantic segmentation tasks, which measures model performance by calculating the overlap between the segmented region and the ground truth annotation. The closer these metrics are to 1, the better the model's performance.

In these equations, TP represents the number of pixels correctly classified as "trail" in astronomical images; TN represents the number of pixels correctly classified as "non-trail"; FN represents the number of "trail" pixels incorrectly classified as "non-trail"; and FP represents the number of "non-trail" pixels incorrectly classified as "trail".

## 4 RESULTS AND DISCUSSIONS

In this section, we applied ASA-U-Net to Mephisto images collected since November 2022, 172 FITS images files were collected as the source of the dataset. Based on the model results obtained from the training and validation process, we saved the best model for testing on the test dataset, with the following test results.

4.1 Comparison of Detection Results

Figure 11, 12 and 13 presents a comparative analysis of detection results from ASA-U-Net and the baseline U-Net on artificial satellite tracks across a variety of astronomical backgrounds. The results indicate that ASA-U-Net's detection capabilities are superior, particularly in scenarios involving faint tracks. Furthermore, ASA-U-Net is more proficient at identifying challenging features, such as celestial bodies that intersect the track path, which leads to fewer missed detections and a higher success rate on these segments. In summary, ASA-U-Net demonstrates a significant reduction in both false positives and false negatives relative to the original U-Net. It maintains robust performance across datasets from different telescope channels and more accurately identifies the true positions of satellite tracks, even when they are difficult to detect.

(a)

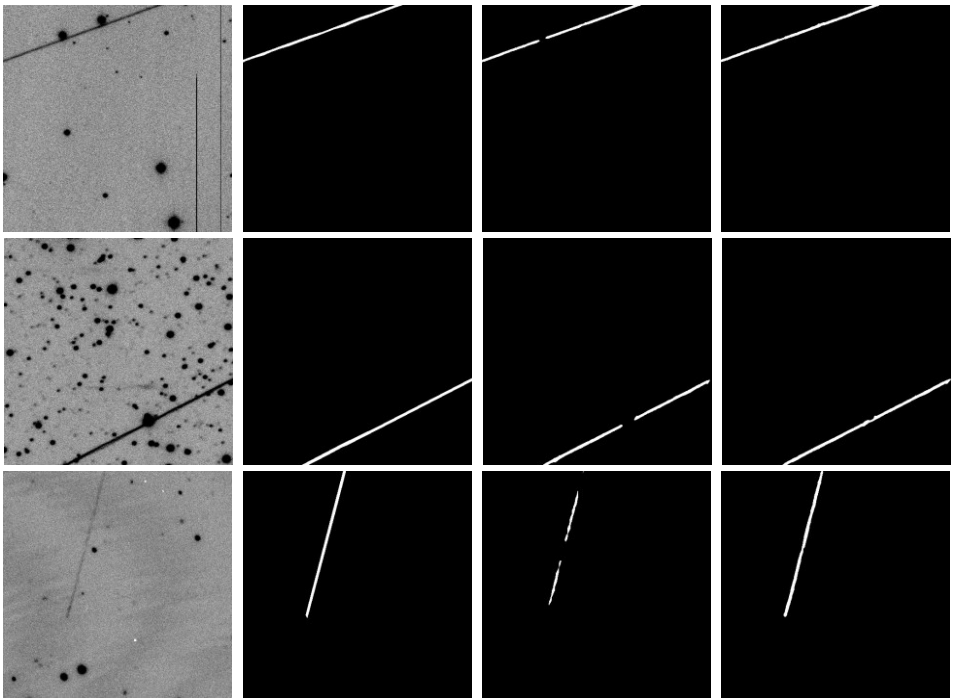

(b)

(c)

Fig.11: The (a), (b), and (c) illustrate prediction results on samples from the red-channel camera dataset. Each set shows, from left to right: the original image, the ground truth label, the U-Net prediction, and the ASA-U-Net models prediction.

## 4.2 Experimental Comparison with other models

(a)

Table 1, 2 and 3 presents the metric data for semantic segmentation experimental results of different deep learning models across different channel cameras. The experimental results show that the ASA-U-Net model overall exhibits the best semantic segmentation performance. Specifically, for the satellite trail category in datasets from different channels, the performance is as follows: In the red channel camera dataset, the IoU of the ASA-U-Net model improved by 1.94% compared to the U-Net model, Dice improved by 1.22%, and Recall increased by 2.22%; in the yellow channel camera dataset, the IoU improved by 0.51%, Dice improved by 0.38%, and Precision increased by 1.4%; in the blue channel camera dataset, which is crucial for astronomical photometry, the IoU improved by 7.7%, Dice improved by 9.99%, Precision increased by 0.92%, and Recall improved by 15.13% compared to the U-Net model. Fig. 14 shows the confusion matrices of the ASA-U-Net model on satellite trail datasets from different channel cameras, where the values and percentages on the main diagonal are high.The experimental results indicate that our proposed ASAU-Net model performs well in semantic segmentation of satellite trails, with fewer false positives and false negatives compared to the U-Net model. Moreover, it does not require line detection algorithms such as Hough, avoiding complex processing workflows. This demonstrates that the model has high recognition accuracy and can meet the task requirements of preventing artificial satellite trail light pollution during astronomical photometry tasks.

(a)

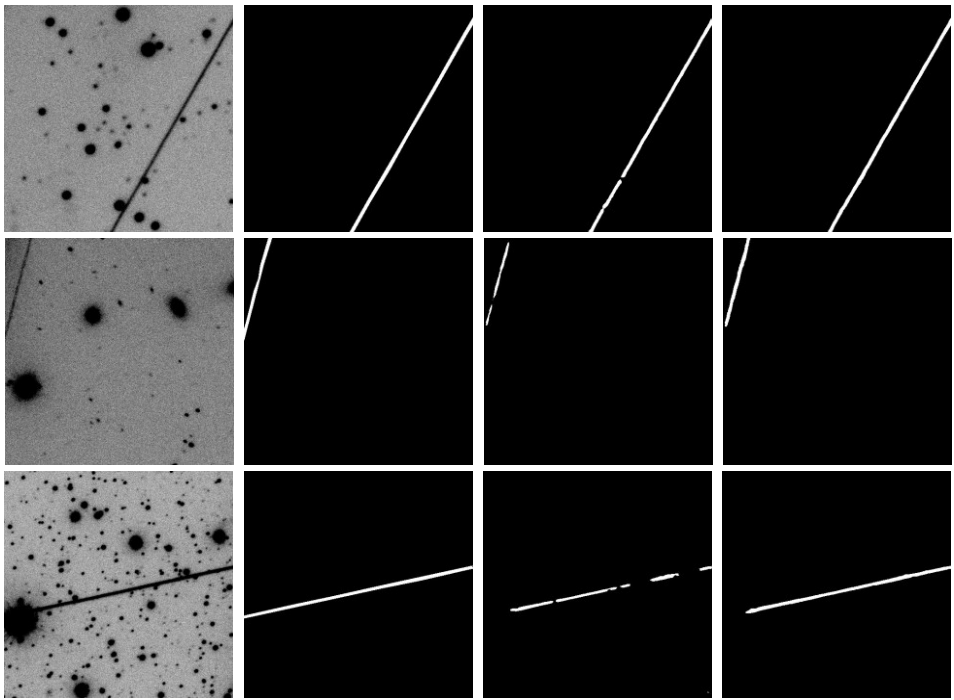

(b)

(c)

Fig.12: The (a), (b), and (c) illustrate prediction results on samples from the yellow-channel camera dataset. Each set shows, from left to right: the original image, the ground truth label, the U-Net prediction, and the ASA-U-Net models prediction.

Table 1: Comparison of evaluation metrics for different deep learning models on the satellite trail dataset from the red channel camera.

| Network | MPA/% | Precision/% | Recall/% | IoU/% | Dice/% |
|---|---|---|---|---|---|
| U-Net | 94.56 | 93.74 | 88.43 | 83.10 | 90.55 |
| ASA-U-Net(Ours) | 95.74 | 93.52 | 90.65 | 85.04 | 91.77 |

(a)

Table 2: Comparison of evaluation metrics for different deep learning models on the satellite trail dataset from the yellow channel camera.

| Network | MPA/% | Precision/% | Recall/% | IoU/% | Dice/% |
| --- | --- | --- | --- | --- | --- |
| U-Net | 95.90 | 92.16 | 92.21 | 85.99 | 91.96 |
| ASA-U-Net(Ours) | 95.95 | 93.56 | 91.71 | 86.50 | 92.34 |

Table 3: Comparison of evaluation metrics for different deep learning models on the satellite trail dataset from the blue channel camera.

| Network | MPA/% | Precision/% | Recall/% | IoU/% | Dice/% |
| --- | --- | --- | --- | --- | --- |
| U-Net | 88.50 | 85.99 | 73.96 | 68.53 | 76.62 |
| ASA-U-Net(Ours) | 92.74 | 86.91 | 89.09 | 76.23 | 86.61 |

4.3 Ablation Experiments

To further validate the effectiveness of the modules used in this paper for enhancing the performance of the

U-Net model, we conducted ablation experiments on the ASPP module and the channel attention mech-

(a)

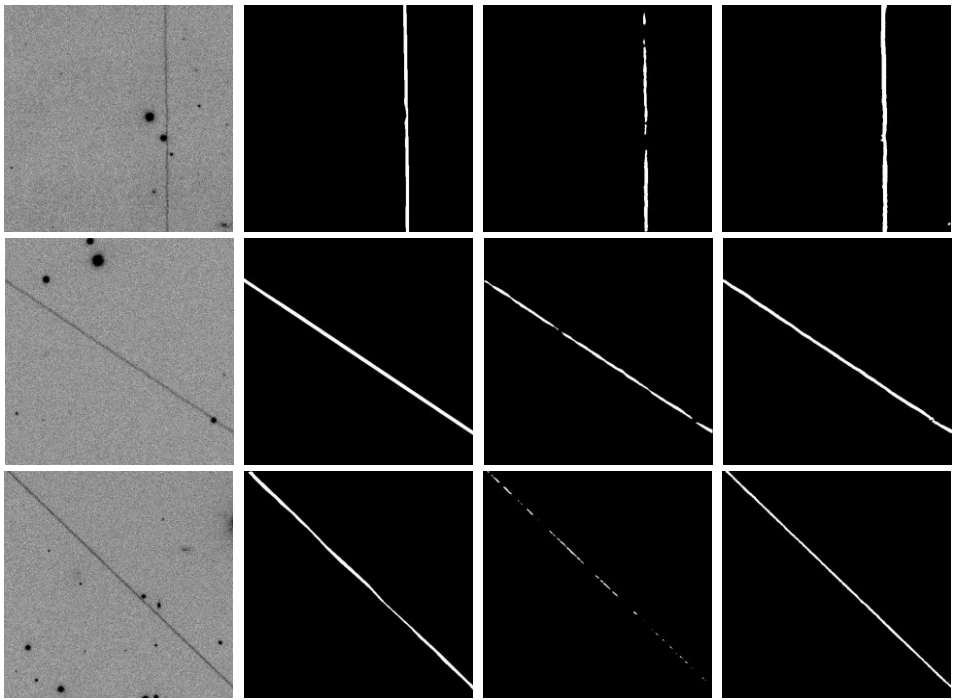

(b)

(c)

Fig.13: The (a), (b), and (c) illustrate prediction results on samples from the blue-channel camera dataset. Each set shows, from left to right: the original image, the ground truth label, the U-Net prediction, and the ASA-U-Net models prediction.

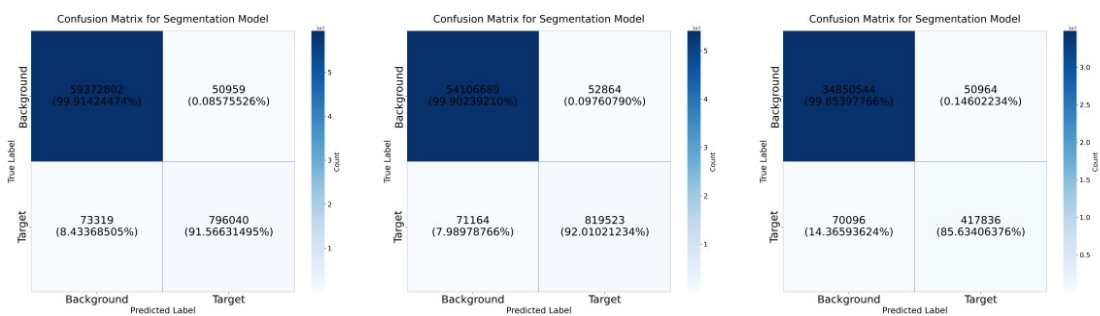

(a)

(a)　　　　　　　　　　　(b)　　　　　　　　　　　(c)

Fig.14: Confusion matrices obtained by the ASA-U-Net model on different test sets; (a): Confusion matrix on the satellite trail test set from the red channel camera. (b): Confusion matrix on the satellite trail test set from the yellow channel camera. (c): Confusion matrix on the satellite trail test set from the blue channel camera.

anism within our ASA-U-Net model. These experiments were designed to compare the impact of each module on the U-Net's performance. The ablation study was carried out on an image dataset captured by the three-channel camera of the Mephisto telescope, with the hyperparameter settings and operating environment during model training remaining consistent with those previously described. We continue to use evaluation metrics such as Mean Pixel Accuracy (MPA), Precision, Recall, Intersection over Union (IoU), and the Dice coefficient.

By analyzing the data from the ablation experiments, we can observe the effects of integrating the ASPP and channel attention modules into the U-Net model. Taking the results from Table 6—which details the ablation study on the blue-channel camera's image dataset where satellite tracks are characteristically faint—we can illustrate the improvements. With hyperparameters and other model structures held constant, the baseline U-Net model achieved a Mean Pixel Accuracy of 88.50%, a Precision of 85.99%, a Recall of 73.96%, an IoU of 68.53%, and a Dice coefficient of 76.62% on this dataset.

Table 4: Ablation experiment results on the satellite trail image dataset from the red channel camera.

| U-Net | ASPP | SAM | MPA/% | Precision/% | Recall/% | IoU/% | Dice/% |
|---|---|---|---|---|---|---|---|
| ✓ |   |   | 94.56 | 93.74 | 88.43 | 83.10 | 90.55 |
| ✓ | ✓ |   | 95.57 | 92.30 | 90.82 | 84.08 | 91.20 |
| ✓ |   | ✓ | 95.70 | 93.01 | 90.88 | 84.83 | 91.64 |
| ✓ | ✓ | ✓ | 95.74 | 93.52 | 90.65 | 85.04 | 91.77 |

Table 5: Ablation experiment results on the satellite trail image dataset from the yellow channel camera.

| U-Net | ASPP | SAM | MPA/% | Precision/% | Recall/% | IoU/% | Dice/% |
|---|---|---|---|---|---|---|---|
| ✓ |   |   | 95.90 | 92.16 | 92.21 | 85.99 | 91.96 |
| ✓ | ✓ |   | 95.51 | 92.24 | 91.21 | 85.06 | 91.49 |
| ✓ |   | ✓ | 94.43 | 93.67 | 88.85 | 84.22 | 90.81 |
| ✓ | ✓ | ✓ | 95.95 | 93.56 | 91.71 | 86.50 | 92.34 |

Table 6: Ablation experiment results on the satellite trail image dataset from the blue channel camera.

| U-Net | ASPP | SAM | MPA/% | Precision/% | Recall/% | IoU/% | Dice/% |
|---|---|---|---|---|---|---|---|
| ✓ |   |   | 88.50 | 85.99 | 73.96 | 68.53 | 76.62 |
| ✓ | ✓ |   | 91.78 | 87.61 | 81.53 | 74.47 | 82.89 |
| ✓ |   | ✓ | 91.10 | 85.03 | 79.91 | 72.58 | 80.68 |
| ✓ | ✓ | ✓ | 92.74 | 86.91 | 89.09 | 76.23 | 86.61 |

To enhance the model's ability to extract target information and to improve its capability to recognize feature channels beneficial to the current task, we first replaced the standard double convolution feature extraction module in the U-Net with our Enhanced DoubleConv (EDC) module. This change resulted in a 2.6% increase in Mean Pixel Accuracy, a 5.95% increase in Recall, a 4.05% increase in IoU, and a 4.06% increase in the Dice coefficient.Subsequently, to improve the model's capacity for extracting features at different scales, we introduced an ASPP module as the bottleneck and added a channel attention mechanism. This led to a further increase of 1.64% in Mean Pixel Accuracy, 1.88% in Precision, 9.18% in Recall, 3.65% in IoU, and 5.93% in the Dice coefficient. With the combination of these two modules, the improved model achieved the best performance.

5 CONCLUSIONS

Based on the characteristics of artificial satellite trails in astronomical images and the demand for highprecision astronomical survey photometry, this paper proposes the ASA-U-Net model, an advancement of U-Net, for detecting satellite trails. This method combines the U-Net architecture with an Atrous Spatial Pyramid Pooling (ASPP) module, a channel attention mechanism, and residual connections to identify satellite trails in images captured by ground-based telescopes. Using data from the Mephisto survey telescope, we constructed three types of image datasets that include challenging conditions such as Fringing effects, dense celestial populations, and faint trails. The experimental results demonstrate the model's effectiveness in identifying and marking satellite trails, achieving high precision, recall, and Dice coefficients, which confirms its capability to effectively recognize satellite trails across different wavebands and complex backgrounds.

Importantly, the model and methodology developed in this research can be integrated into telescope data processing systems. Considering the geometric and morphological features of satellite trails, we used labelme software to perform precise manual annotations, creating a semantic segmentation dataset of astronomical images containing satellite trails. We also completed a series of tasks including image cropping and data augmentation. This approach allows for the efficient extraction of satellite trail features without requiring cumbersome preprocessing by specialists, enabling precise detection. By providing our design and experimental methods, we strive to address the severe challenges that satellite trails pose to astronomical data processing and to improve the quality of various astronomical data. However, the ASA-U-Net model still has areas for improvement. For example, while it has effectively enhanced recall and precision, there remains room for improvement in detection speed and model capacity.

Future work will focus on establishing image datasets with a richer diversity of object categories, incorporating other types of interference that affect astronomical data analysis to further enhance the model's generalization capabilities. Subsequently, the model will be integrated into the data processing pipeline of the Mephisto survey telescope at Yunnan University, and the methodology will be further refined after a trial period.

Acknowledgements This work was funded by the National Natural Science Foundation of China (NSFC) under No.11080922.

References

Andreoni I., et al., 2024, Rubin ToO 2024: Envisioning the Vera C. Rubin Observatory LSST Target of Opportunity program (arXiv:2411.04793), https://arxiv.org/abs/2411.04793 2

Bertin E., Arnouts S., 1996, aaps, 117, 393 2

Buslaev A., Iglovikov V. I., Khvedchenya E., Parinov A., Druzhinin M., Kalinin A. A., 2020, Information, 11 9

Chambers K. C., et al., 2019, The Pan-STARRS1 Surveys (arXiv:1612.05560), https://arxiv.org/abs/1612.05560 2

Chen L.-C., Papandreou G., Schroff F., Adam H., 2017, Rethinking Atrous Convolution for Semantic Image Segmentation (arXiv:1706.05587), https://arxiv.org/abs/1706.05587 4, 10

Dawson W. A., Schneider M. D., Kamath C., 2016, Blind Detection of Ultra-faint Streaks with a Maximum Likelihood Method (arXiv:1609.07158), https://arxiv.org/abs/1609.07158 3

Hassanein A. S., Mohammad S., Sameer M., Ragab M. E., 2015, A Survey on Hough Transform, Theory, Techniques and Applications (arXiv:1502.02160), https://arxiv.org/abs/1502.02160 3

He K., Zhang X., Ren S., Sun J., 2015, Deep Residual Learning for Image Recognition (arXiv:1512.03385), https://arxiv.org/abs/1512.03385 10

Howell S. B., 2006, Handbook of CCD Astronomy, 2 edn. Cambridge Observing Handbooks for Research Astronomers, Cambridge University Press 2

Hu J., Shen L., Albanie S., Sun G., Wu E., 2019, Squeeze-and-Excitation Networks (arXiv:1709.01507), https://arxiv.org/abs/1709.01507 4, 10

Kollo N., Akiyama Y., Peethambaran J., 2023, in 2023 20th Conference on Robots and Vision (CRV). pp 256–264, doi:10.1109/CRV60082.2023.00040 3

Lecun Y., Bottou L., Bengio Y., Haffner P., 1998, Proceedings of the IEEE, 86, 2278 4

Loshchilov I., Hutter F., 2019, Decoupled Weight Decay Regularization (arXiv:1711.05101), https://arxiv.org/abs/1711.05101 13

Meher S. K., Panda G., 2021, European Physical Journal Special Topics, 230, 2285 3

Mroz P., et al., 2022, ́ The Astrophysical Journal Letters, 924, L30 2

Nir G., Zackay B., Ofek E. O., 2018, The Astronomical Journal, 156, 229 3

Paillassa M., Bertin E., Bouy H., 2018, in Di Matteo P., Billebaud F., Herpin F., Lagarde N., Marquette J. B., Robin A., Venot O., eds, SF2A-2018: Proceedings of the Annual meeting of the French Society of Astronomy and Astrophysics. p. Di 4

Paillassa M., Bertin E., Bouy H., 2020, aap, 634, A48 4

Ronneberger O., Fischer P., Brox T., 2015, U-Net: Convolutional Networks for Biomedical Image Segmentation (arXiv:1505.04597), https://arxiv.org/abs/1505.04597 4, 10

Stark D. V., Grogin N., Ryon J., Lucas R., 2022, Improved Identification of Satellite Trails in ACS/WFC Imaging Using a Modified Radon Transform, Instrument Science Report ACS 2022-8, 25 pages 3

Stoppa F., Groot P. J., Stuik R., Vreeswijk P., Bloemen S., Pieterse D. L. A., Woudt P. A., 2024, aap, 692, A199 2

Tiwari A., 2022, in 2022 Fourth International Conference on Transdisciplinary AI (TransAI). pp 30–31, doi:10.1109/TransAI54797.2022.00011 4

Turin G., 1960, IRE Transactions on Information Theory, 6, 311 3

Walker C., et al., 2020, Bulletin of the AAS, 52 2

Waszczak A., et al., 2017, pasp, 129, 034402 3

York D. G., et al., 2000, AJ, 120, 1579 2

Yuan X., et al., 2020, in Marshall H. K., Spyromilio J., Usuda T., eds, Society of Photo-Optical Instrumentation Engineers (SPIE) Conference Series Vol. 11445, Ground-based and Airborne Telescopes VIII. p. 114457M, doi:10.1117/12.2562334 4